\documentclass[prl,twocolumn,showpacs]{revtex4}
\usepackage[dvips]{graphicx}
\usepackage{latexsym}
\usepackage{bm}
\begin{document}
\newcommand{\fig}[2]{\includegraphics[width=#1]{#2}}
\newcommand{\pprl}{Phys. Rev. Lett. \ } 
\newcommand{\pprb}{Phys. Rev. {B}} 
\newcommand{\be}{\begin{equation}}
\newcommand{\ee}{\end{equation}}
\newcommand{\bea}{\begin{eqnarray}} 
\newcommand{\eea}{\end{eqnarray}}
\newcommand{\br}{{\bf{r}}} 
\newcommand{\bbr}{{\bf{R}}} 
\newcommand{\bx}{{\bf{x}}}
\newcommand{\bj}{{\bf{J}}}
\newcommand{\bq}{{\bf{q}}}
\newcommand{\sperp}{{\sigma_v}}
\newcommand{\spara}{{\sigma_v}}
\newcommand{\pll}{\parallel} 
\newcommand{\upa}{\uparrow} 
\newcommand{\dna}{\downarrow}
\newcommand{\rhohat}{{{\hat\rho_s}}}
\newcommand{\zhat}{{{\hat z}}}
\newcommand{\svc}{{\sigma_v^c}}
\newcommand{\fmn}{{f_{\mu\nu}}}
\newcommand{\pmu}{{\partial_\mu}}
\newcommand{\pnu}{{\partial_\nu}}
\newcommand{\plambda}{{\partial_\lambda}}
\newcommand{\bamu}{{\bar a_\mu}}

\title{Vortices, tunneling and deconfinement in 
 bilayer quantum Hall excitonic superfluid}

\author{Ziqiang Wang}
\affiliation{Department of Physics, Boston College, Chestnut Hill, MA 02467}

\date{\today}

\begin{abstract}
The physics of vortices, instantons and deconfinement 
is studied for layered superfluids
in connection to bilayer quantum Hall systems at filling fraction
$\nu=1$. We develop an effective gauge theory taking
into account both vortices and instantons induced by interlayer tunneling. 
The renormalization group flow of the gauge charge and the instanton
fugacity shows that the coupling of the gauge field to vortex matter
produces a continuous transition between the 
confining phase of free instantons and condensed 
vortices and a deconfined gapless superfluid where magnetic charges
are bound into dipoles. 
The interlayer tunneling conductance and the layer imbalance
induced inhomogeneous exciton condensate 
are discussed in connection to experiments.

\typeout{polish abstract}
\end{abstract}

\pacs{73.43.Jn, 73.43.Nq, 73.21.-b, 64.60.Ak}

\maketitle

When the total number of electrons in a double-layer electron system 
equals the degeneracy of the lowest spin-resolved Landau 
level, a stable phase of matter emerges in which an electron in one layer
binds a hole in the other to form an interlayer phase coherent
exciton condensate \cite{physicstoday,book}. 
Such a state is expected to display transport properties similar
to a neutral superfluid (SF) in the layer antisymmetric
channel \cite{fertig,wenzee,ezawa}. Recent experiments,
tunneling \cite{spielman}, Coulomb drag \cite{kellogg}, 
and counterflow \cite{counterflow}, have
evidenced the development of such an excitonic SF at
small layer separations. Yet the same experiments observed
properties that are unconventional of a two-dimensional SF.
The usual vortex-unbinding Kosterlitz-Thouless (KT) transition 
has not been seen; the zero-bias tunneling peak, $\delta$-function
like in a Josephson analogy, is large but finite; and the counterflow 
resistance, zero in the case of a SF, is small but nonzero at
very low temperatures. These findings suggest that the SF
response is limited by the dissipation of quantum 
vortices induced by disorder.
\cite{physicstoday,sbf,fertigvortex,huse}.
Indeed, numerical diagonalizations find that moderate disorder generates
a gauge-glass phase where the SF stiffness is nonzero
only at zero-temperature \cite{shengbalentswang}.

In addition to vortices, another important issue is whether the 
{\it spontaneous} interlayer coherence is stable against interlayer 
tunneling, i.e.
an exchange-enhanced \cite{jorg} symmetric-antisymmetric band 
splitting $\Delta_{\rm SAS}$.
The tunneling events create space-time singularities, i.e. instantons, 
in the counterflow current \cite{wenzee}. 
In analogy to compact quantum
electrodynamics in (2+1) dimensions (QED$_3$), where the U(1)
gauge field is always confining and massive due to
instantons \cite{polyakov}, one may conclude intuitively that the gapless
SF mode, and thus the exciton condensate is destroyed by tunneling. 

The above analogy to a pure gauge theory is oversimplified because of the
coupling to vortex matter.
In this paper, we investigate the interplay between the
vortices and instantons and the emergence of a deconfined
exciton condensate. The subject of (de)confinement when matter fields are 
present is also central to the gauge theory formulation of spin liquids
and underdoped high-T$_c$ cuprates 
\cite{nagaosalee,senthil}. Using the strong-coupling
duality approach, we describe the neutral SF
in terms of vortices coupled to a gauge field.
Instantons induced by tunneling are
magnetic monopoles/antimonopoles with an interaction
mediated by the gauge field.
Taking into account the contributions from both instantons and
vortices with Ohmic dissipation, we obtain the renormalization group (RG) 
flow of the gauge charge and the instanton fugacity. 
We find a fixed point in the RG flow at
a critical vortex conductivity $\svc$. For $\sigma_v < \svc$,
the instantons are in a plasma phase capable of screening; the gapless
mode is destroyed and the gauge field is confining.
This is a $\nu=1$ integer quantum Hall state driven 
by an arbitrarily small interlayer tunneling. The deconfined exciton
condensate emerges for $\sigma_v > \svc$, where
the instanton fugacity flows to zero and the monopoles/antimonopoles bind into
magnetic dipoles in a dielectric phase. We present a perturbative
calculation of the tunneling current under vortex dissipation.
The layer-imbalance is shown to lead to an inhomogeneous SF with spatial 
density modulations.

We begin with the action for the neutral SF mode,
\be
S={1\over2}\rho_s\int d^3r [\partial_\mu\theta(\br)]^2,
\label{s}
\ee
where $\br=(x,y;\tau)$, $\partial_\mu=(\partial_x,\partial_y;\partial_\tau)$,
and $\theta$ is the phase of the excitonic order parameter with
stiffness $\rho_s$. For convenience, we set the
superfluid velocity $v_s$ to unity. The current-density
in the isotropic space-time is 
$J_\mu=\rho_s\partial_\mu\theta$. Eq.(\ref{s}) is
the (2+1)D XY-model in the continuum limit. To account for
the periodicity of $\theta$, we go to the
vortex picture by a duality transformation 
\cite{duality}. Introducing a Hubbard-Stratonovich (HS) field $J_\mu$,
\be
S=\int d^3r\left({1\over2\rho_s}J_\mu^2+iJ_\mu\left[\pmu\theta_0(\br)
+\pmu\theta_{\rm v}(\br)\right]\right).
\label{s1}
\ee
Here we have decomposed $\theta$ into the sum of a
single-valued $\theta_0$ and a multivalued vortex part $\theta_{\rm v}$.
The vortex 3-current is $j_\mu^v=(1/2\pi)\epsilon_{\mu\nu\lambda}
\pnu\plambda\theta_{\rm v}$. Integrating over $\theta_0$ enforces the 
conservation of the boson 3-current, $\pmu J_\mu=0$, i.e.
charge continuity in the {\it absence} of tunneling. 
The latter can be implemented explicitly by writing
$
J_\mu=(1/2\pi)\epsilon_{\mu\nu\lambda}
\pnu a_\lambda,
$
where $a_\mu$ is a {\it noncompact} U(1) gauge field.
The dual action of the SF follows,
\be
S=\int d^3r\left({1\over8\pi^2\rho_s}f_{\mu\nu}^2+i j_\mu^v a_\mu\right),
\label{duals}
\ee
with $\fmn=\pmu a_\nu-\pnu a_\mu$. This is the Maxwell action of a
gauge field minimally coupled to the vortex matter carrying unit 
gauge charge $e_g=\sqrt{2\pi^2\rho_s}$. The vortex current is 
conserved, i.e. $\pmu j_\mu^v=0$, in the {\it absence} of tunneling.
Vortices have closed-loop
world lines. The gapless gauge field
mediates a logarithmic vorticity interaction. Consequently,
there is a KT transition due to the proliferation of the vortices.
In quantum Hall bilayers, vortices are merons carrying half of an
electric charge in addition to vorticity \cite{moonyang}. 
The lowest energy charge excitations are meron pairs of
like charge and opposite vorticity.

As electrons tunnel, meron pairs are excited 
and the counterflow current $J_\mu$ is no longer conserved. 
The continuity equation becomes 
$\pmu J_\mu=\pm2\delta(x)\delta(y)\delta(\tau)$ following a tunneling
event at $\br=0$. Such sources and drains of charge
implies 
$
\int d^3r\pmu J_\mu={1\over2\pi}\int d^3r\epsilon_{\mu\nu\lambda}
\pnu a_\lambda=\pm2,
$
i.e. the emergence of magnetic monopoles/antimonopoles in 
the space-time configuration of $a_\mu$ \cite{wenzee}. 
Thus, although $a_\mu$ starts off as a noncompact gauge field, 
tunneling introduces singularities, i.e.
instantons, that arise spontaneously in a compact U(1) gauge theory.
Writing $a_\mu=a_\mu+a_\mu^{\rm inst}$, where $a_\mu$ is the smooth
part of the gauge field, we have the magnetic Gauss's law
$\epsilon_{\mu\nu\lambda}\pmu\pnu a^{\rm inst}_\lambda=4\pi\rho_M$,
where $\rho_M=\sum_a q_a\delta(\br-\br_a)$ is the instanton/monopole density
and $q_a=\pm1$ are the magnetic charges.
The magnetic charges experience the $1/r$-interaction
from the Maxwell term in Eq.(\ref{duals}) and have the action
\cite{polyakov,wenzee},
\be
S_{\rm inst}={1\over2}{1\over8\pi^2\rho_s}\sum_{a\ne b}
q_a{4\pi\over\vert \br_a-\br_b\vert}q_b+m_c\sum_a q_a^2,
\label{sinst}
\ee
where $m_c$ is the action cost for nucleating an isolated monopole 
core. The magnetic charge $e_m=1/2e_g$ is identified from the first term. 
Decouple the monopole interactions via a HS field $\phi$, 
and sum over the trajectories of monopoles, the instanton
partition function is obtained \cite{polyakov} and the total
path integral of the bilayer is
\bea
Z&=&\int{\cal D}[a,j^v,\phi]
e^{-\int d^3r ({\cal L}_v[a,j^v]+{\cal L}_{\rm inst}[\phi])},
\label{z} \\
{\cal L}_v&=&{1\over8\pi^2\rho_s}f_{\mu\nu}^2+i j_\mu^v a_\mu,
\label{lvortex} \\
{\cal L}_{\rm inst}&=&{1\over2}\rho_s(\pmu\phi)^2-z\cos\phi.
\label{linst}
\eea
The instanton part, Eq.(\ref{linst}), is the
sine-Gordon action with $z=e^{-m_c}$ the instanton fugacity.
Since it represents the probability
of finding a monopole per unit space-time volume, $z$ is directly
related to the tunneling rate, $z\simeq N_0\Delta_{SAS}$
where $N_0=1/2\pi\ell_B^2$ is the electron density. 
Eqs.(\ref{z}-\ref{linst}) describe the system of 
gauge field and instantons coupled to vortex matter.
Without the matter fields,
Polyakov showed that instantons drive the compact QED$_3$
to the confined phase where the pure gauge field
acquires a mass \cite{polyakov}.
This amounts to a relevant fugacity $z$ in the (2+1)D sine-Gordon model
that grows under RG transformations, causing the proliferation of instantons.
The monopoles are in a plasma phase capable of screening
and the gauge charges are confined.

The question is what happens when vortices are present.
Due to instantons, the vortices are no longer
conserved since they are created and annihilated at the monopoles.
The vortex configuration thus
contains both closed loops and open segments (strings) connecting
a monopole and an antimonopole \cite{savit}.
For a nonzero fugacity $z$, there is a nonzero string
tension $T_s\sim \sqrt {z\rho_s}$. The vortices are linearly confined and
large vortex loops are interrupted by instantons.
This is the confinement phase where
monopoles are free and the vortex string condensation destroys
the SF coherence. The existence of a gapless exciton condensate is thus tied 
to the
deconfinement of the gauge field, which is only possible if
the instanton fugacity vanishes. When $z=0$, it costs an infinite amount of
energy to create a free monopole. The monopoles and antimonopoles are thus
bound into magnetic dipoles and magnetic charges are confined
in an insulating dielectric phase. Since the string tension $T_s$ 
goes to zero in this case, the vortex segments disappear. 
The low energy physics is then governed by that
of a SF with finite vortex loops.

We next show that the coupling to dissipative vortex matter indeed
stabilizes a deconfined phase and materializes a quantum deconfinement
transition.
At the {\it classical} level, a related issue
of an $XY$-model in a symmetry-breaking field was recently 
revisited and the deconfinement of thermally excited vortices 
was found \cite{fertigdeconfine}.
It is important to emphasize what we have at hand is a system of vortices
coupled to both the smooth and the singular part of the gauge field.
As a consequence, the electric and magnetic charges are
determined by the same $\rho_s$ in Eqs.(\ref{lvortex}) and
(\ref{linst}). The RG flow equation for $\rho_s$
has therefore two contributions; one
from instantons in the sine-Gordon action Eq.(\ref{linst}),
which is quite standard, and the other from the vortex-gauge field
coupling in the instanton-free sector.
In the presence of disorder,
vortices move along the direction of the Magnus and Lorentz forces, 
as well as drift in the direction of the superflow current. 
This degrades the temporal order of the SF
and introduces dissipation in the transport
channels under both parallel and counterflow geometry.

To study the effects of dissipation due to mobile vortices
in the instanton-free sector, we examine the self-energy corrections 
to the gauge field to quadratic order,
\be
\Delta S_2={1\over2}\sum_{\bq,\omega}
\Pi_{\mu\nu}(\bq,\omega)a_\mu(\bq,\omega)a_\nu(-\bq,-\omega),
\label{pijj}
\ee
where $\Pi_{\mu\nu}=\langle j_\mu^v j_\nu^v\rangle$ is the vortex
current-current correlation function. Since the
vortices are mobile at extremely low temperatures,
$\Pi_{\mu\nu}$ must have a form consistent with quantum dissipation.
Note that one cannot formally ``integrate out'' the 
quantum dissipative vortices because they correspond to
gapless low energy degrees of freedom.
Instead, Eq.(\ref{pijj}) should be understood in the sense
of the RG, i.e. at appropriate momentum and frequency scales.
Moreover, the RG procedure must be carried out simultaneously
in the instanton sector because there is not an adiabatic limit justifying
a separation of energy scales. The gauge symmetry implies that the gauge
fields in $\Delta S_2$ must appear in the form of $f_{\mu\nu}^2$ in
the long-wavelength and low-energy limit, leading
to a renormalization of $\rho_s$ or the gauge charge.
The absence of nonlinearity in the tunneling and
counterflow transport \cite{spielman,counterflow}
suggests that the vortices are described by Ohmic dissipation.
This is in fact consistent with the current-current correlation
of the $XY$-model above the KT transition \cite{halperin79}. 
We thus take $\Pi_{\mu\nu}$ to be of the diffusive form
and denote the vortex
conductivity by $\sigma_v$ in unit of 
$e_v^2/\hbar$ with $e_v$ the vortex charge. We have, in the Coulomb gauge
${\vec \nabla}\cdot{\vec a}=0$,
\be
\Delta S_2={1\over2}\sum_{\bq,\omega}\left[
{\vert\omega\vert\sperp\over q^2}f_{xy}^2(\bq,\omega)
+{\spara\over\vert\omega\vert}f_{i,\tau}^2(\bq,\omega)\right],
\label{ds2}
\ee
where the index $i=(x,y)$. The singular self-energy
corrections in the long-wavelength 
and low-energy limit indicate that the gauge field acquires
an anomalous dimension. Introducing the dimensionless coupling
$\rhohat=b^{-1}Z_a\rho_s$, where $b$ is an isotropic 3-momentum scale
and $Z_a$ is the renormalization constant of the gauge field,
Eq.(\ref{ds2}) leads to
$Z_a=1-4\pi^2 \eta b^{-1}\sperp\rho_s$. Here,
$\eta$ is a dimensionless constant relating $q=b/\eta$.
It is now straightforward to obtain the RG flow
$\beta_\rho= d\rhohat/d\ell$ with $\ell=-\ln b$. Combined with
the standard RG contributions from the sine-Gordon action in Eq.(\ref{linst})
under the same cutoff procedure, we arrive at the complete flow equations
that account for both instantons and vortices,
\bea
{d\rhohat\over d\ell}&=&\rhohat-4\pi^2 \eta \sperp\rhohat^2
+8{\zhat^2\over\rhohat}
\label{betarhos}
\\
{d\zhat\over d\ell}&=&\left(3-{1\over4\pi^2\rhohat}\right)\zhat,
\label{betaz}
\eea
where $\zhat$ is the dimensionless fugacity and the $\zhat^2$ term 
comes from screening by monopole-antimonopole pairs.

The RG flow in the $\rhohat$-$\zhat$ plane determines the possible phases
of the system. Without mobile vortices, i.e.
for $\sperp=0$, both $\zhat$ and $\rhohat$ flow to large values.
The monopoles stay in the single plasma phase and the gauge field is confining.
In the presence of vortex dissipation, however,
a deconfined phase emerges for sufficiently large $\sperp$.
Indeed, the flow equations (\ref{betarhos}) and (\ref{betaz})
support a nontrivial fixed point at
$\rhohat^*={1\over4\pi^2\eta\sperp}$ and $\zhat^*=0$,
provided that $\sperp>\svc=3/\eta$.
In this case, the scaling dimension of $z$ becomes larger than the
space-time dimension, forcing a finite fugacity (tunneling rate)
to flow to zero while the phase stiffness to a finite value. This fixed point 
is infrared stable and represents a deconfined phase
where the monopoles and antimonopoles are bound into dielectric dipoles.
The critical value $\svc$ marks a confinement-deconfinement transition.
These are our main results. The deconfinement transition 
is from the single-particle quantum Hall fluid driven by interlayer 
tunneling to the spontaneous phase coherent exciton condensate.
A direct transition from the latter to
the incoherent compressible state is also possible since it has been shown
that a dissipative normal fluid can suppress the interlayer 
tunneling at low temperatures \cite{wang}.

We next discuss the interlayer tunneling conductance in the 
presence of vortex dissipation. 
The perturbative calculation \cite{sbf} of the tunneling current 
density $J(t)=(e/\hbar)z\langle\sin(\theta-\omega t)\rangle$ is
valid in the deconfined phase where 
the instanton fugacity $z$ is irrelevant in the RG. 
Under a bias voltage $V$, the current density
is given by \cite{sbf},
\be
J(V)=C_0\!\int_0^\infty\! dt\!\int\! d^2R \sin({eV\over\hbar}t)
{\theta(v_st-R)\over\sqrt{v_s^2t^2-R^2}}e^{-W(\bbr,t)},
\label{jv0}
\ee
where $C_0=ev_sz^2/2\pi\hbar\rho_s$. The Debye-Waller factor $e^{-W(\bbr,t)}
=\langle e^{i\theta(\bbr,t)}e^{-i\theta(0,0)}\rangle
=\langle e^{i\int d^3r j_\mu\partial_\mu\theta}\rangle$,
where $j_\mu(\bx,\tau;\bbr,t)=\delta_{\mu,0}[\delta(\bx)\theta(\tau)
-\delta(\bx-\bbr)\theta(\tau-t)]$ is the current along the
path connecting the tunneling/instanton events at $(0,0)$ and $(\bbr,t)$.
The quantum average is with respect to the dual action in
Eq.~(\ref{duals}). In the presence of $j_\mu$, the
saddle-point (SP) can be solved by writing $a_\mu=\delta a_\mu + \bamu$,
$\bamu=\partial_\mu\theta$. The SP equation is
$\partial_\tau \rho_v+{\vec\nabla}\cdot{\vec j_v}=\partial_\mu j_\mu
=\rho_M$, which connects the divergence of the vortex current to the
instanton density $\rho_M=\delta(\bx)\delta(\tau)
-\delta(\bx-\bbr)\delta(\tau-t)$. From vortex diffusion, we have
${\vec j_v}=D({\vec\nabla}-\gamma_H {\hat z}
\times{\vec\nabla})\rho_v$ where $D$ is the diffusion constant
and $\gamma_H$ is the Hall angle which
does not affect charge spreading in the bulk. Thus, the
tunneling induced vortex density is
$
\rho_v(\bq,\omega_n;\bbr,t)
=\Gamma(\bq,\omega_n;\bbr,t)/(Dq^2+\vert\omega_n\vert),
$
where $\Gamma=1-e^{i(\bq\cdot\bbr-\omega_n\tau)}$. Separating
$j_\mu^v=\delta j_\mu^v+{\bar j_\mu^v}$ with 
${\bar j_\mu^v}=\rho_v\delta_{\mu,0}$, the Debye-Waller factor
becomes 
$e^{-W(\bbr,t)}=\langle e^{i\int d^3r \rho_v (\br;\bbr,t)\delta a_0(\br)}
\rangle$ evaluated with the Lagrangian density
${\cal L}=(1/2e_g^2)(\partial_\mu \delta a_0)^2+
i\delta\rho_v \delta a_0$ in the Coulomb gauge.
Thus $W$ depends on the interaction mediated by the 
longitudinal gauge field $\delta a_0$,
\be
W(\bbr,t)={T\over2}\sum_{q,\omega_n}V_{\rm vort}(\bq,\omega_n)
{\vert\Gamma(\bq,\omega_n;\bbr,t)\vert^2\over(Dq^2+\vert\omega_n\vert)^2},
\label{w}
\ee
where $V_{\rm vort}=(q^2/e_g^2)+\sigma_v q^2/\hbar(Dq^2+\vert\omega_n\vert)$ 
is the screened vortex interaction.
In the perturbative regime, $t\ll t_p=\sqrt{\rho_s/zv_s^2}$, screening
due to the slow diffusive motion of the vortices
can be ignored for weak dissipation (small
vortex conductivity), since $t\ll t_v= \hbar/e_g^2\sigma_v$ is always 
satisfied. Indeed, $t_v/t_p
=\hbar\omega_J/2\pi^2\rho_s\sigma_v\gg1$ can be verified explicitly 
using the estimates \cite{book}, $\omega_J\sim 10^{10}/{\rm s}$, 
$\rho_s\sim 0.1$ K, and $\sigma_v\sim 3\times10^{-4}$ from the 
counterflow resistivity \cite{counterflow}. We find, in this regime, 
$
W(\bbr,t)= {t\over\tau_\varphi}+{R^2\over2\xi_\varphi},
$
where $\tau_\varphi=8\hbar/\pi\rho_s$ is the coherence time for recombination
of the vortex-antivortex pair and
$\xi_\varphi=\sqrt{\pi D\tau_\varphi/4}$ is the coherence length.
Performing the integral in Eq.~(\ref{jv0}), we obtain the tunneling
conductance
\be
{dJ(V)\over dV}={e J_0}{\hbar/\tau_\varphi\over (eV)^2+(\hbar/\tau_\varphi)^2},
\label{djdv}
\ee
where $J_0=ez^2\xi_\varphi^2/\hbar\rho_s$.
The vortex dissipation destroys the quasi-long-range temporal order of the 
SF phase and removes the Josephson singularity. The tunneling 
conductance shows a Lorentzian peak around zero bias with a half-width 
$\hbar/\tau_\varphi=\pi\rho_s/8\sim 3 \ \mu$eV. This value is expected to
be renormalized downward by quantum fluctuation corrections to the
phase stiffness in the Hartree-Fock theory \cite{book,moonyang}.
These results are consistent with the tunneling experiments \cite{spielman}. 
The full behavior of the tunneling current will be discussed elsewhere.

To summarize, instantons and deconfinement
are important for understanding layered superfluids
in the presence of interlayer tunneling. For quantum Hall
bilayers, the interlayer coherent exciton condensate
is realized in the deconfined phase stabilized by mobile vortices with
Ohmic dissipation.
The transition between the gapless exciton condensate
and the single-particle quantum Hall state is described by the 
deconfinement transition.
We conclude with a discussion on imbalanced bilayers created by
an applied bias voltage ($\mu$) coupled to the difference of the
carrier density. 
Retracing the steps leading to the dual action shows that
the effect of imbalance is to add a term $-\mu({\vec \nabla}\times {\vec a})$
to Eqs.(\ref{duals}) and (\ref{lvortex}). 
The net gauge flux ${\vec\nabla}\times {\vec a}$
represents the {\it finite} boson density-imbalance and the
chemical potential $\mu$ acts as an applied magnetic field.
Increasing layer-imbalance induces more vortices and enhances
dissipation (an increase in $\rho_{xx}$).
In analogy to type II superconductors,
at a critical value $\mu_c$, vortices will condense. 
In the {\it absence} of tunneling/instantons, 
the gauge flux forms an Abrikosov lattice.
This state is a Wigner crystal of bosons
and the transition would be in the universality
class of a field-tuned superconductor-insulator transition \cite{jinwu}.
Indeed, Hartree-Fock theory predicts an incipient charge density 
wave order triggered by the softening of the magneto-roton minimum
\cite{jorg}.
This conventional picture, however, breaks down due to
instantons induced by interlayer tunneling. The reason being, once vortices 
condense, the gauge field $a_\mu$ in Eq.(\ref{lvortex}) acquires a mass
due to the Anderson-Higgs mechanism. The instanton fugacity $z$ 
in Eq.(\ref{linst}) becomes relevant. Thus, vortex condensation 
implies $\phi=0$ and a gain of the Josephson energy 
$E_J=-z\langle\cos\phi\rangle$. In a flux lattice, the phase
$\phi$ winds around each vortex by $2\pi$ and cannot gain
the Josephson energy. The Wigner crystal of bosons is thus unstable
due to instantons. The competition between flux lattice and vortex
condensation leads naturally to an inhomogeneous state as a 
compromise. The system breaks up into puddles of vortex condensate
where $E_J\ne0$ separated by regions of uncondensed vortices where $E_J=0$. 
The existence of such an inhomogeneous excitonic condensate
is consistent with the observation that the interlayer coherence 
is robust against layer imbalance 
\cite{sawada,tutucimbalance,jimimbalance}.
With increasing imbalance,
macroscopic phase coherence eventually establishes across the
puddles via vortex tunneling and a transition into an inhomogeneous,
compressible normal state takes place. This corresponds to
the layer-imbalance driven transition observed 
experimentally \cite{jimimbalance}.

The author thanks J.P. Eisenstein, S.M. Girvin, A.H. MacDonald,
and X.G. Wen for discussions, and Aspen Center for Physics for hospitality.
This work is supported in part by DOE grant No. DE-FG02-99ER45747 
and ACS grant No. 39498-AC5M.


\begin{thebibliography}{99}
\bibitem{physicstoday} J.P. Eisenstein and A.H. MacDonald,
cond-mat/0404113.
\bibitem{book}For reviews, see S.M. Girvin and A.H. MacDonald,
in {\sl Perspectives in Quantum Hall Effect}, edited by S. Das Sarma
and A. Pinczuk (Wiley, New York, 1997).
\bibitem{fertig} 
H. Fertig, \pprb {\bf40}, 1087 (1989).
\bibitem{wenzee}
X.G. Wen and A. Zee, \pprl {\bf 69}, 1811 (1992);
\bibitem{ezawa}
Z.F. Ezawa and A. Iwazaki, \pprb {\bf48}, 15189 (1993).
\pprb {\bf47}, 2265 (1993).
\bibitem{spielman}
I.B. Spielman {\it et al.}, \pprl {\bf 84}, 5808 (2000); 
{\bf87}, 036803 (2001).
\bibitem{kellogg}
M. Kellogg {\it et. al.,} \pprl {\bf90}, 246801 (2003).
\bibitem{counterflow}
M. Kellogg {\it et. al.,} \pprl {\bf93}, 036801 (2004);
E. Tutuc {\it et. al.,} \pprl {\bf93}, 036802 (2004).
\bibitem{sbf}  
A. Stern {\it et al.}, \pprl {\bf 86}, 1829 (2001);
L. Balents and L. Radzihovsky, {\it ibid.} {\bf 86}, 1825 (2001);
M. Fogler and F. Wilczek, {\it ibid.} {\bf 86}, 1833 (2001).
\bibitem{fertigvortex}
H.A. Fertig and J.P. Straley, {\pprl} {\bf91}, 046806 (2003).
\bibitem{huse}
D.A. Huse, cond-mat/0407452 (2004).
\bibitem{shengbalentswang}
D. Sheng, L. Balents, and Z. Wang
\pprl {\bf91}, 116802 (2003).
\bibitem{jorg}
Y.N. Joglekar and A.H. MacDonald \pprb{\bf65}, 235319 (2002).
\bibitem{polyakov}
A.M. Polyakov, Nucl. Phys. B{\bf120}, 429 (1977); 
{\sl Gauge field and Strings} (Harwood Academic, London 1987).
\bibitem{nagaosalee}
N. Nagaosa and P.A. Lee, \pprb{\bf61}, 9166 (2000).
\bibitem{senthil}
M. Hermele {\it et al.,} \pprb{\bf70}, 214437 (2004).
\bibitem{duality}
M.P.A. Fisher and D.H. Lee, \pprb{\bf39}, 2756 (1989).
\bibitem{moonyang}
K. Moon {\it et al.}, Phys. Rev. B {\bf 51}, 5138 (1995); 
K. Yang {\it et al.}, {\it ibid.} {\bf 54}, 11644 (1996).
\bibitem{savit} 
M.B. Einhorn and R. Savit, Phys. Rev. D{\bf17}, 2583 (1978);
{\bf19}, 1198 (1979).
\bibitem{fertigdeconfine}
H.A. Fertig, \pprl{\bf89}, 035703 (2002).
\bibitem{halperin79}
B.I. Halperin and D.R. Nelson, J. Low Temp. Phys. {\bf36}, 599 (1979).
\bibitem{wang} Z. Wang, \pprl{\bf92}, 136803 (2004).
\bibitem{jinwu}
J.W. Ye, cond-mat/0407088 (2004).
\bibitem{sawada}
A. Sawada {\it et al.,} \pprl{\bf80}, 4534 (1998).
\bibitem{tutucimbalance}
E. Tutuc {\it et al.,} \pprl{\bf91}, 076802 (2003).
\bibitem{jimimbalance}
I.B. Spielman {\it et al.,} cond-mat/0406067 (2004).
\end{thebibliography}
\end{document}